\documentclass[12pt,tightenlines,eqsecnum,floats,aps,amsmath,amssymb,nofootinbib,superscriptaddress,showpacs,prd]{revtex4}

\usepackage{dcolumn}
\usepackage{bm}
\usepackage[spanish,english]{babel}
\usepackage{amsfonts}
\usepackage{amssymb}
\usepackage{graphicx}
\usepackage[latin1]{inputenc}

\newcommand{\be}{\begin{equation}}
\newcommand{\ee}{\end{equation}}
\newcommand{\bea}{\begin{eqnarray}}
\newcommand{\eea}{\end{eqnarray}}

\begin{document}

\title{{\bf Stellar Structure Equations in Extended Palatini Gravity}}

\author{Gonzalo J. Olmo}\email{gonzalo.olmo@csic.es}

\affiliation{Departamento de F\'{i}sica Te\'{o}rica and
IFIC,  Universidad de Valencia-CSIC, Facultad de F\'{i}sica, C/ Dr. Moliner 50,
Burjassot-46100, Valencia, Spain.}

\author{H\`{e}lios Sanchis-Alepuz}
\affiliation{ Fachbereich Theoretische Physik, Institut f\"{u}r Physik, Karl-Franzens-Universit\"{a}t Graz,
Universit\"{a}tsplatz 5, A-8010 Graz, Austria}
\affiliation{Departamento de F\'{i}sica Te\'{o}rica and
IFIC,  Universidad de Valencia-CSIC, Facultad de F\'{i}sica, C/ Dr. Moliner 50,
Burjassot-46100, Valencia, Spain.}

\author{Swapnil Tripathi}\email{swapnil.tripathi@uwc.edu}
\affiliation{University of Wisconsin-Washington County, CSEPA,
400 University Drive  West Bend, WI 53095, USA}

\begin{abstract}
We consider static spherically symmetric stellar configurations in 
Palatini theories of gravity in which the Lagrangian is an unspecified 
function of the form $f(R,R_{\mu\nu}R^{\mu\nu})$. We obtain the 
Tolman-Oppenheimer-Volkov equations corresponding to this class of theories 
and show that they recover those of $f(R)$ theories and General Relativity 
in the appropriate limits. We show that the exterior vacuum solutions are of Schwarzschild-de Sitter type and comment on the possible expected modifications, as compared to GR, of the interior solutions. 
\end{abstract}


\maketitle

\vspace{1cm}

\section{Introduction}\label{sec:introduction}

One of the most challenging open questions that Physics faces nowadays is that of explaining why the Universe has the structures that we observe  and how they came into existence. The current establishment states that some yet-to-be-determined source of dark energy should be responsible for the observed speed up in the cosmic expansion rate \cite{reviews}. Additionally, it assumes the existence of huge amounts of dark matter, which must have been necessary to increase the rate of growth of structures out of the highly homogeneous initial configuration observed in the cosmic microwave background radiation \cite{Weinberg2008} (see also \cite{sanders}). The effects of the dark matter component should be observable today in the dynamical and kinematical properties of stars in galaxies and of galaxies in clusters.  \\
Though the dark matter and dark energy models are able to successfully fit different data sets, they also suffer from some limitations that motivate the search for alternative explanations (see for instance \cite{Famaey:2011kh}). In this sense, scenarios in which the gravitational interaction is modified at large scales have been proposed to address the same phenomenology as the dark matter and dark energy models. This has led in the last years to investigate theoretically and observationally the effects of possible deviations of the gravitational dynamics in the Newtonian and Einsteinian regimes (see, for instance, the review articles \cite{MyReview,Famaey:2011kh, Review-f(R)}). \\

Though the existence of dark energy has very little effect, if any, on the structure of stars, the presence of dark matter or of modified dynamics can have a much more dramatic impact which, in some cases, could lead to observational effects. In fact, it has been suggested that dark matter could play a non-trivial role in the mechanisms of energy generation and transport in stars, which could be used to set bounds on the type and properties (decay rates) of the hypothetical particles making up the dark sources \cite{DMStars}. The potential effects of modified gravitational dynamics are more difficult to estimate, however, because one must derive the precise set of equations that govern the internal structure of fluids in equilibrium and then consider specific models of interest, such as relativistic stars or compact objects, which unavoidably requires the use of numerical methods. Despite these difficulties, the study of the effects of alternative equations on the structure of compact objects, stellar formation and evolution, and peculiar objects (instability strips, protostars, etc) has attracted some attention in the context of  modified theories of gravity of different types \cite{DeLaurentis:2012st, starsBeyondGR,Olmo08b,starsPalatini}. Given the increasingly high accuracy and resolution of currently available observational techniques (see for instance \cite{Chamel:2008ca,Abbott:2008fx}) and the discovery of rare objects which defy the standard model of stellar formation and evolution \cite{magnetars}, the study of  stellar properties and new solutions in extensions of GR could provide new insights to interpret observations and offer new avenues to test the strong field regime of gravitation and, therefore, help determine or set limits to potential corrections to Einstein's equations.  \\

Motivated by all these reasons, in this work we study how the structure equations of perfect fluids in hydrostatic equilibrium, the so-called  Tolman-Oppenheimer-Volkoff (TOV) equations, are modified with respect to those found in General Relativity (GR) when the gravitational Lagrangian is extended to become a function of the Ricci scalar and the Ricci squared scalar, i.e., $f(R, Q)$, with $Q=g^{\mu\alpha}g^{\nu\beta}R_{\mu\nu}R_{\alpha\beta}=R_{\mu\nu}R^{\mu\nu}$. Though the usual metric formulation\footnote{The metric formulation assumes that the connection is constrained {\it a priori} to be given by the Christoffel symbols of the metric.} of this type of theories  is known to have ghosts and other dynamical perturbative instabilities due to the higher-order character of the field equations, we  consider its Palatini (or metric-affine or first-order)  formulation \cite{MyReview}, which is free from those drawbacks, as will be explained in detail below. The family of Palatini $f(R, Q)$ theories have been studied in the recent literature specially in scenarios involving high-energy physics, such as early-time cosmology \cite{BO2010}, black holes \cite{O&R2012}, and applications to quantum gravity phenomenology \cite{Olmo2011QG}, and also in the context of the cosmic speedup \cite{Bauer:2010bu, LBM07, ABF04}. In the usual (metric) formulation theories of that kind are also well known (see, for instance, \cite{B&O1983,Anderson} for some early references, and \cite{Review-f(R)} for more recent works). \\ 

The Palatini formulation of $f(R,Q)$ theories leads to a set of second-order equations for the metric that exactly recover GR in vacuum. In regions containing sources, such as in the interior of stars or fluids in general, the dynamics is modified due to nonlinear matter terms induced by the form of the Lagrangian. These nonlinearities arise due to the nontrivial role played by the matter in the determination of the affine connection, which is assumed to be independent of the metric (Palatini formalism), i.e., it is not constrained a priori to be given by the Christoffel symbols of the metric. As a result, material systems whose gravitational dynamics is governed by a Lagrangian different from that of Hilbert-Einstein, $f(R,Q)=R$, could admit equilibrium configurations different from those found in GR, which could have observable consequences. \\
Although Palatini $f(R,Q)$ theories do not introduce higher-order differential equations, they involve elaborated algebraic manipulations (required to solve for the connection) that lead to nontrivial modifications of the dynamics. For this reason, in this work we  focus on the derivation of the corresponding TOV equations and on putting them in a suitable form that facilitates their use in numerical applications, which will be considered elsewhere. We also compare the resulting expressions with those obtained in the literature for Palatini $f(R)$ theories. \\

The content of this paper is organized as follows. In Section \ref{sec:act_field_eqs} we review the formulation of $f(R,Q)$ theories in the Palatini formalism. We then derive explicit expressions for the case of perfect fluids in subsection \ref{sec:PerFluids} and for simple models of the $f(R,Q)$ action in subsection \ref{sec:model}. The stellar structure equations in those models are derived in Section \ref{sec:stell_str_eqs}, where we also check that GR is recovered when the appropriate limit is taken. These results and its potential applications are discussed in Section \ref{sec:summary}.

\section{Action and field equations}\label{sec:act_field_eqs} 

We define Palatini $f(R,Q)$ theories as follows
\begin{equation}\label{eq:action}
S[g,\Gamma,\psi_m]=\frac{1}{2\kappa^2}\int d^4x \sqrt{-g}f(R,Q) +S_m[g,\psi_m],
\end{equation}
where $\kappa^2\equiv 8\pi G$,  $S_m[g,\psi_m]$ represents the matter action, $g_{\alpha\beta}$ is the space-time metric, $R=g^{\mu\nu}R_{\mu\nu}$, $Q=g^{\mu\alpha}g^{\nu\beta}R_{\mu\nu}R_{\alpha\beta}$, $R_{\mu\nu}={R^\rho}_{\mu\rho\nu}$, and
${R^\alpha}_{\beta\mu\nu}=\partial_{\mu}
\Gamma^{\alpha}_{\nu\beta}-\partial_{\nu}
\Gamma^{\alpha}_{\mu\beta}+\Gamma^{\alpha}_{\mu\lambda}\Gamma^{\lambda}_{\nu\beta}-\Gamma^{\alpha}_{\nu\lambda}\Gamma^{\lambda}_{\mu\beta}$. 
Variation of (\ref{eq:action}) with respect to metric and connection leads to the following equations \cite{OSAT09}
\begin{eqnarray}
f_R R_{\mu\nu}-\frac{f}{2}g_{\mu\nu}+2f_QR_{\mu\alpha}{R^\alpha}_\nu &=& \kappa^2 T_{\mu\nu}\label{eq:met-varX}\\
\nabla_{\beta}\left[\sqrt{-g}\left(f_R g^{\mu\nu}+2f_Q R^{\mu\nu}\right)\right]&=&0  \ ,
 \label{eq:con-varX}
\end{eqnarray}
were we have used the short-hand notation $f_X\equiv \partial_X f$. For simplicity, in the above derivation we have assumed a symmetric Ricci tensor, $R_{\mu\nu}=R_{\nu\mu}$, and vanishing torsion (for more details on the effects of relaxing these conditions, see \cite{O&R2012x}).  As shown in \cite{OSAT09}, the connection equation (\ref{eq:con-varX}) can be solved introducing a rank-two tensor (or auxiliary metric) $h_{\mu\nu}$ such that $\sqrt{-g}\left(f_R g^{\mu\nu}+2f_Q R^{\mu\nu}\right)=\sqrt{-h}h^{\mu\nu}$, which turns  (\ref{eq:con-varX}) into $\nabla_\beta \left[\sqrt{-h}h^{\mu\nu}\right]=0$ and implies that the connection can be expressed as the Levi-Civita connection of $h_{\mu\nu}$. Therefore, the Ricci tensor of the independent connection is equivalent to the Ricci tensor of the metric $h_{\mu\nu}$, which from now on we denote $R_{\mu\nu}(h)$.  With simple algebraic manipulations, one finds that the relation between $h_{\mu\nu}$ and $g_{\mu\nu}$ can be expressed as follows
\begin{equation} \label{eq:h-g}
h_{\mu\nu}=\sqrt{\det\Sigma} {[\Sigma^{-1}]_\mu}^\alpha g_{\alpha\nu}  \ \text{   } , \text{   } \ h^{\mu\nu}=\frac{g^{\mu\alpha}{{\Sigma_\alpha} ^\nu}}{\sqrt{\det\Sigma}} \ , 
\end{equation}
 where we have defined the matrix ${\Sigma_\alpha} ^\nu\equiv \left(f_R \delta_\alpha^\nu +2f_Q {B_\alpha}^\nu\right)$ and $ {B_\alpha}^\nu=R_{\alpha\beta}(h)g^{\beta\nu}$. It is important to note (see section \ref{sec:model} below for details) that the matrix  ${B_\alpha}^\nu$ is a function of the components of the stress-energy tensor. This implies that $R={B_\alpha}^\alpha$ and $Q={B_\alpha}^\nu{B_\nu}^\alpha$ are functions of the matter. According to this, the physical metric $g_{\mu\nu}$ and the auxiliary metric $h_{\mu\nu}$  are related by a matter-dependent deformation given by the matrix ${\Sigma_\alpha} ^\nu$. This deformation becomes a conformal factor in the particular case of Palatini $f(R)$ theories (where $f_Q\to 0$), as is well known \cite{MyReview}.\\ 
 With these relations and definitions, the field equations (\ref{eq:met-varX}) for the metric $h_{\mu\nu}$ can be written in compact form as ${B_\mu}^\alpha {\Sigma_\alpha}^\nu=\frac{f}{2}\delta_\mu^\nu+\kappa^2{T_\mu}^\nu$, and using the relation ${B_\mu}^\alpha {\Sigma_\alpha}^\nu=\sqrt{\det\Sigma} R_{\mu\alpha}(h) h^{\alpha\nu}$ we obtain 
\begin{equation}\label{eq:field-eqs}
{R_\mu}^\nu(h)=\frac{1}{\sqrt{\det\Sigma}}\left(\frac{f}{2}\delta_\mu^\nu+\kappa^2{T_\mu}^\nu\right) \ .
\end{equation}
This form of the metric field equations will be used in what follows to derive the TOV stellar structure equations. However, before that we need to determine the explicit form of the matrix ${\Sigma_\alpha} ^\nu$. That task is carried out in the next subsection.

\subsection{ $f(R,Q)$ theories with a perfect fluid }\label{sec:PerFluids}

The explicit form of the matrix $\Sigma$ that relates the metrics $h_{\mu\nu}$ and $g_{\mu\nu}$ can only be found once all the sources that make up $T_{\mu\nu}$ have been specified. In our discussion we will just consider a perfect fluid or a sum of non-interacting perfect fluids such that 
\begin{equation}\label{eq:Tmn}
T_{\mu\nu}=(\rho+P)u_\mu u_\nu+P g_{\mu\nu}
\end{equation}
with $\rho=\sum_i \rho_i$ and $P=\sum_i P_i$. In order to find an expression for $\Sigma$, we first rewrite (\ref{eq:met-varX}) using matrix notation as
\begin{equation}
2f_Q\hat{B}^2+f_R \hat{B}-\frac{f}{2}\hat{I} = \kappa^2 \hat{T} \label{eq:met-varRQ2} \ .
\end{equation}
Using (\ref{eq:Tmn}) this equation can be rewritten as follows
\begin{equation}
2f_Q\left(\hat B+\frac{f_R}{4f_Q}\hat I\right)^2=\left(\kappa^2 P+\frac{f}{2}+\frac{f_R^2}{8f_Q}\right)\hat I+\kappa^2(\rho+P)u_\mu u^\mu\label{eq:met-varRQ3} \ .
\end{equation} 
Denoting $\lambda^2\equiv \left(\kappa^2P+\frac{f}{2}+\frac{f_R^2}{8f_Q}\right)$ and making explicit the matrix representation, (\ref{eq:met-varRQ3}) becomes
\begin{equation}
2f_Q\left(\hat B+\frac{f_R}{4f_Q}\hat I\right)^2=\begin{pmatrix}
 \lambda^2-\kappa^2(\rho+P) & {0}  \\
{0} & \lambda^2 \hat{I}_{3X3}
\end{pmatrix}   \label{eq:met-varRQ4} \ ,
\end{equation} 
where $\hat{I}_{3X3}$ denotes 3-dimensional identity matrix. Since the right-hand side of (\ref{eq:met-varRQ4}) is a diagonal matrix, it is immediate to compute its square root, which leads to 
\begin{equation}
\sqrt{2f_Q}\left(\hat B+\frac{f_R}{4f_Q}\hat I\right)=\begin{pmatrix}
 s_1\sqrt{\lambda^2-\kappa^2(\rho+P)} & {0}  \\
{0} & \lambda \hat{S}_{3X3}
\end{pmatrix}   \label{eq:met-varRQ5} \ ,
\end{equation} 
where $s_1$ denotes a sign, which can be positive or negative, and $\hat{S}_{3X3}$ denotes a $3X3$ diagonal matrix with elements $\{s_i=\pm 1\}$. For consistency of the theory in the limit $f_Q\to 0$, we must have $s_1=1$ and $\hat{S}_{3X3}=\hat{I}_{3X3}$. This result allows to express the matrix ${\Sigma_\mu}^\nu$ as follows
\begin{equation}
\hat \Sigma=\begin{pmatrix}
 \sigma_1 & {0}  \\
{0} & \sigma_2 \hat{I}_{3X3}
\end{pmatrix}   \label{eq:Sigma} \ ,
\end{equation} 
where a hat denotes matrix representation, and $\sigma_1$ and $\sigma_2$ take the form
\begin{eqnarray}
\sigma_1&=&  \frac{f_R}{2}\pm \sqrt{2f_Q}\sqrt{\lambda^2-\kappa^2(\rho+P)}\nonumber\\
\sigma_2&=& \frac{f_R}{2}+\sqrt{2f_Q}\lambda \ . \label{eq:sigmas}
\end{eqnarray}
Note that we have kept the two signs in front of the square root of $\sigma_1$. In order to correctly recover GR at low densities, one must take the positive sign in that equation. However, at high densities the square root may vanish and one may need to take the negative sign branch to guarantee that $\sigma_1$ is continuous and differentiable accross the point where the square root vanishes (this subtlety in the behavior of $\sigma_1$ was first observed in \cite{BO2010}). This technical issue does not arise for $\sigma_2$.  \\
 
\subsection{Workable models: $f(R,Q)=\tilde{f}(R)+\alpha Q$ }\label{sec:model}

So far we have made progress without specifying the form of the Lagrangian $f(R,Q)$. However, in order to find the explicit dependence of $R={B_\mu}^\mu$ and $Q={B_\mu}^\alpha{B_\alpha}^\mu$ with the $\rho$ and $P$ of the fluids, we must choose a Lagrangian explicitly. Restricting the function $f(R,Q)$ to the family 
$f(R,Q)=\tilde{f}(R)+\alpha Q$, we will see that it is possible to find the generic dependence of $Q$ with $\rho$ and $P$, while $R$ is found to depend only on the combination $T=-\rho+3P$ \cite{OSAT09,OSAT09MG}. The reason for this follows from the trace of (\ref{eq:met-varX}) with $g^{\mu\nu}$, which for this family of Lagrangians gives the algebraic relation $R \tilde{f}_R-2\tilde{f}=\kappa^2T$ and implies that $R=R(T)$ (like in Palatini $f(R)$ theories). For these theories, we have that $f_Q=\alpha$, which is a constant. Therefore, from the trace of (\ref{eq:met-varRQ4}) we find
\begin{equation}
\sqrt{2f_Q}\left(R+\frac{f_R}{f_Q}\right)=
\sqrt{\lambda^2-\kappa^2(\rho+P)} +3\lambda\ ,
\end{equation}
which can be cast as
\begin{equation}
\left[\sqrt{2f_Q}\left(R+\frac{f_R}{f_Q}\right)-3\lambda\right]^2=
\lambda^2-\kappa^2(\rho+P)
\end{equation}
After a bit of algebra we find that 
\begin{equation}\label{eq:lambda1}
\lambda= \frac{\sqrt{2f_Q}}{8}\left[3\left(R+\frac{f_R}{f_Q}\right)\pm\sqrt{\left(R+\frac{f_R}{f_Q}\right)^2-\frac{4\kappa^2(\rho+P)}{f_Q}}\right]
\end{equation}
From this expression and the definition of $\lambda^2$,  we find


\begin{equation} \label{eq:Q}
\alpha Q=-\left(\tilde f+\frac{\tilde f_R^2}{4f_Q}+2\kappa^2P\right)+\frac{{f_Q}}{16}\left[3\left(R+\frac{\tilde f_R}{f_Q}\right)\pm\sqrt{\left(R+\frac{\tilde f_R}{f_Q}\right)^2-\frac{4\kappa^2(\rho+P)}{f_Q}}\right]^2 \ ,
\end{equation}


where $R$, $\tilde{f}$, and $\tilde f_R$ are functions of $T=-\rho+3P$.\\

\section{Stellar Structure equations}\label{sec:stell_str_eqs}

\subsection{Geometric part. Preliminaries. }
Given the field equations in the form (\ref{eq:field-eqs}) and  the matrix $\Sigma$ of a perfect fluid, see (\ref{eq:Sigma}),  we have all the elements to compute the Ricci tensor $R_{\mu\nu}(\Gamma)=R_{\mu\nu}(h)$,  which represents the left-hand side of the field equations (\ref{eq:field-eqs}). Choosing the diagonal metric $g_{\mu\nu}$ as $g_{\mu\nu}\to(-A(r)e^{2\psi(r)},1/A(r),r^2,r^2\sin^2\theta)$, the corresponding $h_{\mu\nu}$ diagonal elements are 
\begin{eqnarray}
h_{tt}&=&\frac{\sigma_2^2}{\sqrt{\sigma_1 \sigma_2}}g_{tt}\equiv S g_{tt}\\
h_{ij}&=&\sqrt{\sigma_1 \sigma_2}g_{ij}=\Omega g_{ij} \ .
\end{eqnarray}
The inverse metric components are trivially found from these ones. 
The non-zero Christoffel symbols are (obtained by direct computation)\\
\begin{center}
\begin{tabular}{lll}
$\Gamma^t_{tr}= \frac{1}{2}\frac{\partial_r[Sg_{tt}]}{Sg_{tt}]}$ & $ \Gamma^r_{tt}=-\frac{\partial_r[g_{tt}S]}{2\Omega g_{rr}} $  & $\Gamma^r_{\theta\theta}=-\frac{\partial_r[\Omega r^2]}{2\Omega g_{rr}} $ \\
$\Gamma^r_{\phi\phi}= \sin^2\theta \Gamma^r_{\theta\theta}$ & $ \Gamma^r_{rr}= \frac{\partial_r[\Omega g_{rr}]}{2\Omega g_{rr}}$  & $\Gamma^\theta_{r\theta}= \frac{\partial_r[\Omega r^2]}{2\Omega r^2}$ \\
$\Gamma^\theta_{\phi\phi}= -\sin\theta\cos\theta$ & $ \Gamma^\phi_{r\phi}= \Gamma^\theta_{r\theta}$  & $\Gamma^\phi_{\phi\theta}=\frac{\cos\theta}{\sin\theta} $ 
\end{tabular}
\end{center}
For completeness, we expand those coefficients as follows

\begin{center}
\begin{tabular}{lll}
$\Gamma^t_{tr}= \frac{1}{2}\left(\frac{A_r}{A}+2\psi_r+\frac{S_r}{S}\right)$ & $ \Gamma^r_{tt}=-\frac{1}{2}\frac{S g_{tt}}{\Omega g_{rr}}\left(\frac{A_r}{A}+2\psi_r+\frac{S_r}{S}\right) $  & $\Gamma^r_{\theta\theta}=-\frac{r^2A}{2}\left(\frac{\Omega_r}{\Omega}+\frac{2}{r}\right)$ \\
$\Gamma^r_{\phi\phi}= \sin^2\theta \Gamma^r_{\theta\theta}$ & $ \Gamma^r_{rr}= \frac{1}{2}\left(\frac{\Omega_r}{\Omega}-\frac{A_r}{A}\right)$  & $\Gamma^\theta_{r\theta}= \frac{1}{2}\left(\frac{\Omega_r}{\Omega}+\frac{2}{r}\right)$ \\
$\Gamma^\theta_{\phi\phi}= -\sin\theta\cos\theta$ & $ \Gamma^\phi_{r\phi}= \Gamma^\theta_{r\theta}$  & $\Gamma^\phi_{\phi\theta}=\frac{\cos\theta}{\sin\theta} $ 
\end{tabular}
\end{center}

The terms that contribute to the Ricci tensor components are the following:
\begin{eqnarray}
R_{tt}&=& \partial_r\Gamma^r_{tt}+\Gamma^r_{tt}\left(\Gamma^r_{rr}+2\Gamma^\theta_{r\theta}-\Gamma^t_{rt}\right)\\
R_{rr}&=& -\partial_r\left(\Gamma^t_{rt}+2\Gamma^\theta_{r\theta}\right)+\Gamma^r_{rr}\left(\Gamma^t_{rt}+2\Gamma^\theta_{r\theta}\right)-\left((\Gamma^t_{rt})^2+2(\Gamma^\theta_{r\theta})^2)\right) \\
R_{\theta\theta}&=& 1+\partial_r\Gamma^r_{\theta\theta}+\Gamma^r_{\theta\theta}\left(\Gamma^t_{rt}+\Gamma^r_{rr}\right)\\
R_{\phi\phi}&=& \sin^2\theta R_{\theta\theta}
\end{eqnarray}
Inserting the corresponding Christoffel symbols, the result is
\begin{eqnarray}
R_{tt}&=& -\frac{1}{2}\left(\frac{S g_{tt}}{\Omega g_{rr}}\right)\left[\frac{A_{rr}}{A}-\left(\frac{A_r}{A}\right)^2+2\psi_{rr}+\frac{S_{rr}}{S}-\left(\frac{S_r}{S}\right)^2+\right.\nonumber \\
& & \left.\left\{\frac{A_r}{A}+2\psi_r+\frac{S_r}{S}\right\}\left\{\frac{1}{2}\left(\frac{S_r}{S}+\frac{\Omega_r}{\Omega}\right)+\psi_r+\frac{2}{r}+\frac{A_r}{A}\right\}\right]\\
R_{rr}&=& -\frac{1}{2}\left[\frac{A_{rr}}{A}-\left(\frac{A_r}{A}\right)^2+2\psi_{rr}+\frac{S_{rr}}{S}-\left(\frac{S_r}{S}\right)^2+2\left\{\frac{\Omega_{rr}}{\Omega}-\left(\frac{\Omega_r}{\Omega}\right)^2\right\}+\right.\nonumber\\ &+&\left. 
\left\{\frac{A_r}{A}+2\psi_r+\frac{S_r}{S}\right\}\left\{\frac{A_r}{A}+\psi_r-\frac{1}{2}\left(\frac{\Omega_r}{\Omega}-\frac{S_r}{S}\right)\right\}\right]-\frac{A_r}{2A}\left(\frac{2}{r}+\frac{\Omega_r}{\Omega}\right)-\frac{\Omega_r}{r\Omega} \\
R_{\theta\theta}&=& 1-A(1+r\psi_r)-rA_r-\frac{r^2 A}{2}\left[\frac{\Omega_r}{\Omega}\left(\psi_r+\frac{4}{r}+\frac{A_r}{A}\right)+\frac{1}{2}\left(\frac{\Omega_r}{\Omega}+\frac{2}{r}\right)\left(\frac{S_r}{S}-\frac{\Omega_r}{\Omega}\right)+\frac{\Omega_{rr}}{\Omega}\right]
\end{eqnarray}

\subsection{Metric field equations}

We now focus on the equations of motion (\ref{eq:field-eqs}). Using the notation ${R_\mu}^\nu={\tau_\mu}^\nu$, we find the following useful relations
\begin{eqnarray}
\frac{\Omega}{S}{R_t}^t-{R_r}^r &=& \frac{\Omega}{S}{\tau_t}^t-{\tau_r}^r \\
{R_\theta}^\theta&=&{\tau_\theta}^\theta
\end{eqnarray}
Using the results of above for the Ricci tensor, we find

\begin{eqnarray}
\frac{\Omega}{S}{R_t}^t-{R_r}^r &=& -A\left[\psi_r\left(\frac{2}{r}+\frac{\Omega_r}{\Omega}\right)+\frac{1}{2}\frac{\Omega_r}{\Omega}\left(2\frac{\Omega_r}{\Omega}+\frac{S_r}{S}\right)+\frac{1}{r}\left(\frac{S_r}{S}-\frac{\Omega_r}{\Omega}\right)-\frac{\Omega_{rr}}{\Omega}\right]\\
{R_\theta}^\theta&=& \frac{2M_r}{r^2}-\frac{A\psi_r}{r}-\frac{A}{2}\left[\frac{\Omega_r}{\Omega}\left(\psi_r+\frac{4}{r}+\frac{A_r}{A}\right)+\frac{1}{2}\left(\frac{\Omega_r}{\Omega}+\frac{2}{r}\right)\left(\frac{S_r}{S}-\frac{\Omega_r}{\Omega}\right)+\frac{\Omega_{rr}}{\Omega}\right]
\end{eqnarray}

We can thus write the structure equations as follows:
\begin{eqnarray}\label{eq:psir}
\left(\frac{\Omega_r}{\Omega}+\frac{2}{r}\right)\psi_r &=&\frac{1}{A}\left[ \tau_r^r-\frac{\Omega}{S}\tau^t_t\right]     
-\frac{1}{2}\frac{\Omega_r}{\Omega}\left(2\frac{\Omega_r}{\Omega}+\frac{S_r}{S}\right)-\frac{1}{r}\left(\frac{S_r}{S}-\frac{\Omega_r}{\Omega}\right)+\frac{\Omega_{rr}}{\Omega}
\\
\frac{2M_r}{r^2}&=&{\tau_\theta}^{\theta} +\frac{A}{2}\left[\frac{\Omega_r}{\Omega}\left(\frac{4}{r}+\frac{A_r}{A}\right)+\frac{1}{2}\left(\frac{\Omega_r}{\Omega}+\frac{2}{r}\right)\left(\frac{S_r}{S}-\frac{\Omega_r}{\Omega}\right)+\frac{\Omega_{rr}}{\Omega}\right]+\frac{A\psi_r}{2}\left(\frac{\Omega_r}{\Omega}+\frac{2}{r}\right)
\end{eqnarray}

The second equation can be further simplified using (\ref{eq:psir}) and becomes 
\begin{equation}
\frac{4M_r}{r^2}=3\tau_r^r-\frac{\Omega}{S}\tau_t^t+A\left[\frac{2\Omega_{rr}}{\Omega}+\frac{\Omega_{r}}{\Omega}\left(\frac{A_r}{A}+\frac{4}{r}-\frac{3}{2}\frac{\Omega_{r}}{\Omega}\right)\right] \ ,
\end{equation}

where we have used that for a perfect fluid ${\tau_\theta}^{\theta}={\tau_r}^{r}$.  With a bit more of extra effort, it can be put in its (almost) definitive form as follows
\begin{equation}\label{eq:Mr}
\left(\frac{\Omega_r}{\Omega}+\frac{2}{r}\right)\frac{M_r}{r}=\frac{3\tau_r^r-\frac{\Omega}{S}\tau_t^t}{2}+A\left[\frac{\Omega_{rr}}{\Omega}+\frac{\Omega_{r}}{\Omega}\left(\frac{2r-3M}{r(r-2M)}-\frac{3}{4}\frac{\Omega_{r}}{\Omega}\right)\right]
\end{equation}
This last expression can be directly compared with equation (11) of \cite{Olmo08b}, where $f(R)$ theories were considered. The $f(R)$ limit is recovered taking $\Omega=S=f_R$,  and a wrong factor in \cite{Olmo08b} is corrected. \\

\subsection{Conservation equation}

Let us now focus on the conservation equation $\nabla_\mu T^{\mu\nu}=0$, which takes the form
\begin{equation}\label{eq:conservation}
\frac{dP}{dr}=-(\rho+P)\left(\psi_r+\frac{A_r}{2A}\right)=-(\rho+P)\left(\frac{M}{r^2A}+\psi_r-\frac{M_r}{rA}\right) \ .
\end{equation}
Combining (\ref{eq:psir}) and (\ref{eq:Mr}), we find
\begin{equation}
\left(\frac{\Omega_r}{\Omega}+\frac{2}{r}\right)\left(\psi_r-\frac{M_r}{rA}\right)=-\frac{\tau_r^r+\frac{\Omega}{S}\tau^t_t}{2A}-\frac{\Omega_r}{\Omega}\left[\frac{3}{4}\frac{\Omega_r}{\Omega}+\frac{2r-3M}{r(r-2M)}\right]-\frac{1}{2}\left(\frac{S_r}{S}-\frac{\Omega_r}{\Omega}\right)\left(\frac{\Omega_r}{\Omega}+\frac{2}{r}\right)
\end{equation}
It is important to note that this expression is independent of $\Omega_{rr}$, which guarantees that (\ref{eq:conservation}) will become a second-order algebraic equation for $P_r$, as it happens in the $f(R)$ case and in GR. 
Denoting $\Delta\equiv\left(\frac{\Omega_r}{\Omega}+\frac{2}{r}\right)$, (\ref{eq:conservation}) becomes
\begin{equation}\label{eq:conservation2}
\frac{dP}{dr}=-(\rho+P)\left[\frac{M}{r^2A}-\frac{(\tau_r^r+\frac{\Omega}{S}\tau^t_t)}{2\Delta A}-\frac{1}{\Delta}\frac{\Omega_r}{\Omega}\left(\frac{3}{4}\frac{\Omega_r}{\Omega}+\frac{2r-3M}{r(r-2M)}\right)-\frac{1}{2}\left(\frac{S_r}{S}-\frac{\Omega_r}{\Omega}\right)\right]
\end{equation}
Assuming an equation of state of the form $\rho=\rho(P)$,  we can formally denote $\Omega_r\equiv \Omega_P P_r$ and $S_r\equiv S_P P_r$, where $\Omega_P\equiv \partial_P \Omega$ and $S_P\equiv \partial_P S$. With these definitions, 
(\ref{eq:conservation2}) can be written as 
\begin{equation}\label{eq:conservation3}
C_1+C_2P_r+C_3P_r^2=0 \ ,
\end{equation}
where 
\begin{eqnarray}
C_1&=& \frac{2(\rho+P)}{r^2(r-2M)}\left[M-\left(\tau^r_r+\frac{\Omega}{S}\tau_t^t\right)\frac{r^3}{4}\right]\\
C_2 &=& \frac{2}{r}\left[1-\frac{(\rho+P)}{2}\left(\frac{\Omega_P}{\Omega}+\frac{S_P}{S}\right)\right]\\
C_3 &=& \frac{\Omega_P}{\Omega}\left[1-\frac{(\rho+P)}{2}\left\{\frac{3}{2}\frac{\Omega_P}{\Omega}-\left(\frac{S_P}{S}-\frac{\Omega_P}{\Omega}\right)\right\}\right]
\end{eqnarray}
One can now compare these expressions with those obtained in \cite{Olmo08b} for the $f(R)$ case.  We find two typos, one of which corresponds to propagating the error previously found in (\ref{eq:Mr}), and affects the factor $2$ in front of $C_2$. The second typo appears in the definition of $C_3$, where the factor $1$ that appears in the square bracket is missing in \cite{Olmo08b}.  Using the same notation as in \cite{Olmo08b}, we can express the pressure as follows:
\begin{equation}\label{eq:pressure}
P_r=-\frac{P^{(0)}_r}{[1-\alpha(r)]}\frac{2}{\left[1\pm\sqrt{1-\beta(r)P^{(0)}_r}\right]} \ ,
\end{equation}
where the $+$ sign in front of the square root should be taken to recover the GR limit, and have used the shorthand notation
\begin{eqnarray}
P^{(0)}_r&=& \frac{(\rho+P)}{r(r-2M)}\left[M-\left(\tau_r^r+\frac{\Omega}{S}\tau_t^t\right)\frac{r^3}{4}\right] \\
\alpha(r) &=& \frac{(\rho+P)}{2}\left(\frac{\Omega_P}{\Omega}+\frac{S_P}{S}\right)\\
\beta(r) &=& (2r)\frac{\Omega_P}{\Omega}\left[1-\frac{(\rho+P)}{2}\left(\frac{3}{2}\frac{\Omega_P}{\Omega}-\left\{\frac{\Omega_P}{\Omega}-\frac{S_P}{S}\right\}\right)\right]
\end{eqnarray}


\subsection{Summary of results}
	
We can now write together the three equations that determine the stellar structure in $f(R,R_{\mu\nu}R^{\mu\nu})$ Palatini theories of gravity:

\begin{eqnarray}\label{eq:psir-fin}
\left(\frac{\Omega_r}{\Omega}+\frac{2}{r}\right)\psi_r &=&\frac{1}{A}\left[ \tau_r^r-\frac{\Omega}{S}\tau^t_t\right]     
-\frac{1}{2}\frac{\Omega_r}{\Omega}\left(2\frac{\Omega_r}{\Omega}+\frac{S_r}{S}\right)-\frac{1}{r}\left(\frac{S_r}{S}-\frac{\Omega_r}{\Omega}\right)+\frac{\Omega_{rr}}{\Omega}
\\
\label{eq:Mr-fin}
\left(\frac{\Omega_r}{\Omega}+\frac{2}{r}\right)\frac{M_r}{r}&=&\frac{3\tau_r^r-\frac{\Omega}{S}\tau_t^t}{2}+A\left[\frac{\Omega_{rr}}{\Omega}+\frac{\Omega_{r}}{\Omega}\left(\frac{2r-3M}{r(r-2M)}-\frac{3}{4}\frac{\Omega_{r}}{\Omega}\right)\right]\\
P_r&=&-\frac{P^{(0)}_r}{[1-\alpha(r)]}\frac{2}{\left[1\pm\sqrt{1-\beta(r)P^{(0)}_r}\right]} \label{eq:pressure-fin}
\end{eqnarray}

\subsection{Additional manipulations needed\label{sec:Prr}}

In order to put the equations in suitable form to do numerical calculations, we still need to specify the form of $\Omega_{rr}$. Note that once an equation of state is given, we can express $\Omega_r$ and $S_r$ as $\Omega_P P_r$ and $S_P P_r$, respectively. The term $\Omega_{rr}$ needs some extra manipulations, because it has the form $\Omega_{rr}=\Omega_{PP} P_r^2+\Omega_P P_{rr}$. After some algebra, one obtains the following expression for $P_{rr}$ 
\begin{equation}
\frac{P_{rr}}{P_{r}}=\frac{P^{(0)}_{rr}}{P^{(0)}_{r}}\left[1+\frac{s}{2}\frac{\beta P^{(0)}_{r}}{\sqrt{1-\beta P^{(0)}_{r}}(1+s \sqrt{1-\beta P^{(0)}_{r}})}\right]+\left[\frac{\alpha_r}{1-\alpha}+\frac{s}{2}\frac{\beta_r P^{(0)}_{r}}{\sqrt{1-\beta P^{(0)}_{r}}(1+s \sqrt{1-\beta P^{(0)}_{r}})}\right]
\end{equation}
where $s=\pm1$, $\alpha_r=\alpha_P P_r$, $\beta_r=\beta_P P_r$, and [here we define $\Phi\equiv\left(\tau_r^r+\frac{\Omega}{S}\tau_t^t\right)$]
\begin{equation}
\frac{P^{(0)}_{rr}}{P^{(0)}_{r}}=\left(\frac{1+\rho_P}{\rho+P}\right)P_r-\frac{2(r-M)}{r(r-2M)}-\frac{r^3}{4}\left(\frac{\Phi_P P_r+\frac{3\Phi}{r}}{M-\Phi\frac{r^3}{4} }\right)+M_r\left(\frac{2}{r-2M}+\frac{1}{M-\Phi\frac{r^3}{4}}\right)
\end{equation}
This last equation can be equivalently written as
\begin{equation}
\frac{P^{(0)}_{rr}}{P^{(0)}_{r}}=\left[\left(\frac{1+\rho_P}{\rho+P}\right)-\frac{\Phi_P \frac{r^3}{4}}{\left(M-\Phi\frac{r^3}{4}\right)}\right]P_r-\left(\frac{2(r-M)}{r(r-2M)}+\frac{ \frac{3\Phi r^2}{4}}{M-\Phi\frac{r^3}{4} }\right)+M_r\left(\frac{2}{r-2M}+\frac{1}{M-\Phi\frac{r^3}{4}}\right)
\end{equation}
This shows that $P_{rr}$ can be expressed in terms of $r, \rho(P), P, P_r, M,$ and $M_r$. Therefore, our system of equations only involves first derivatives of the functions $P_r, M_r$, and $\psi_r$.

\subsection{Limit to $f(R)$ and GR }

Equations (\ref{eq:psir-fin}), (\ref{eq:Mr-fin}), and (\ref{eq:pressure-fin}) allow to obtain the TOV equations corresponding to Palatini $f(R)$ theories  by just taking $\Omega=S\to f_R$ and setting $f(R,Q)\to {f}(R)$. One then obtains 
\begin{eqnarray}\label{eq:psir-fin-fR}
\left(\frac{f_{R,r}}{f_{R}}+\frac{2}{r}\right)\psi_r &=&\frac{1}{A}\left[ \tau_r^r-\tau^t_t\right]     
-\frac{3}{2}\left(\frac{f_{R,r}}{f_{R}}\right)^2+\frac{f_{R,rr}}{f_{R}} \nonumber
\\
\label{eq:Mr-fin-fR}
\left(\frac{f_{R,r}}{f_{R}}+\frac{2}{r}\right)\frac{M_r}{r}&=&\frac{3\tau_r^r-\tau_t^t}{2}+A\left[\frac{f_{R,rr}}{f_{R}}+\right.\nonumber \\ & & \left.\frac{f_{R,r}}{f_{R}}\left(\frac{2r-3M}{r(r-2M)}-\frac{3}{4}\frac{f_{R,r}}{f_{R}}\right)\right] \ .
\end{eqnarray}

The expression for the gradient of the pressure is the same as (\ref{eq:pressure-fin}) except for the fact that $P^{(0)}_r$, $\alpha(r)$, and $\beta(r)$ get redefined as follows
\begin{eqnarray}
P^{(0)}_r&=& \frac{(\rho+P)}{r(r-2M)}\left[M-\left(\tau_r^r+\tau_t^t\right)\frac{r^3}{4}\right] \\
\alpha(r) &=& (\rho+P)\frac{f_{R,P}}{f_{R}}\\
\beta(r) &=& (2r)\left(\frac{f_{R,P}}{f_{R}}\right)\left[1-(\rho+P)\frac{3 f_{R,P}}{4f_{R}}\right]
\end{eqnarray}
The limit to GR simply requires to take $f_R\to 1$ and $f(R)=R$ after taking the $f(R)$ limit, which leads to 
\begin{eqnarray}\label{eq:psir-GR}
\frac{2\psi_r}{r} &=&\frac{1}{A}\left[ \tau_r^r-\tau^t_t\right]=\frac{\kappa^2(\rho+P)}{A}    \\
\frac{2M_r}{r^2}&=&\frac{3\tau_r^r-\tau_t^t}{2}= \kappa^2\rho \label{eq:Mr-fin-fR} \\
P_r&=&  - \frac{(\rho+P)}{r(r-2M)}\left[M-\left(\tau_r^r+\tau_t^t\right)\frac{r^3}{4}\right] \nonumber \\ &= &  - \frac{(\rho+P)}{r(r-2M)}\left[M+\frac{\kappa^2P r^3}{2}\right] \ ,
\end{eqnarray}
where $A=1-2M(r)/r$, and in GR ${\tau_r}^r=\kappa^2(\rho-P)/2$ and ${\tau_t}^t=-\kappa^2(\rho+3P)/2$.

\section{Summary and Discussion}\label{sec:summary}

In this work we have derived the structure equations for hydrostatic equilibrium of spherically symmetric systems in a family of Palatini theories of gravity where the Lagrangian is a function of the form $f(R,R_{\mu\nu}R^{\mu\nu})$. Therefore, the TOV equations of GR have been extended to a much wider family of theories of gravity. We have shown that the differential character of the GR equations is retained, since we only have first-order derivatives of the functions $\Psi(r), M(r)$, and $P(r)$, but new nonlinear contributions (specially of pressure terms) appear due to the non-trivial role played by the matter sources in the determination of the connection.  The corresponding limits to the cases of $f(R)$ theories and GR have been explicitly computed and some typos on previous literature have been corrected. \\

We note that with the expression for $P_{rr}$ obtained in Section (\ref{sec:Prr}), one can  further manipulate (\ref{eq:Mr-fin}) to arrange all the $M_r$ terms together on the left-hand side. That is the form of the equations that should be used in numerical computations. However, since that representation does not provide any new physical insight, we have omitted that step here. We also note that the (unique) exterior solution of the structure equations is of the Schwarzschild-de Sitter type. Since outside of the star $\rho$ and $P$ vanish, we find that $\sigma_1=\sigma_2=$constant, which implies that $\Omega=S=\sigma_1$ and ${\tau_r}^r={\tau_t}^t=$constant, and the equations boil down to $\psi_r=0$ and $M_r=(f/4\sigma_1^2)|_{vac}r^2$, where $(f/4\sigma_1^2)|_{vac}$ is evaluated in vacuum and plays the role of an effective cosmological constant. \\

Though interior solutions will be studied in detail elsewhere for different choices of the function $f(R,R_{\mu\nu}R^{\mu\nu})$, let us briefly discuss the potential effects of considering the following family of quadratic models: $f(R,Q)=R+aR^2/R_P+R_{\mu\nu}R^{\mu\nu}/R_P$, for which $R$ behaves exactly like in GR, $R=-\kappa^2T$, and $Q=R_{\mu\nu}R^{\mu\nu}$ is given in (\ref{eq:Q}). Here $R_P$ is assumed to be some high curvature scale, such as the Planck scale. If we take $a=-1/2$, we find\cite{OSAT09,OSAT09MG} 
\begin{equation}\label{eq:Q-1/2}
Q=\frac{3R_P^2}{8}\left[1-\frac{2\kappa^2(\rho+P)}{R_P}+\frac{2\kappa^4(\rho-3P)^2}{3R_P^2}-\sqrt{1-\frac{4\kappa^2(\rho+P)}{R_P}}\right] \ .
\end{equation}
At low energies, this expression recovers the GR limit, $Q\approx \left(3 P^2+\rho ^2\right)+\frac{3 (P+\rho )^3}{2 R_P}+\frac{15 (P+\rho )^4}{4 {R_P}^2}+\ldots$, but at very high energies, positivity of the argument in the square root of (\ref{eq:Q-1/2}) implies that $\kappa^2(\rho+P)\leq {R_P}/{4}$, which clearly shows that the combination $\rho+P$ is bounded from above
regardless of the symmetries or particular configuration of the fluid involved. 
 The interior solutions of this model must be very similar to those of GR except at the innermost regions of extremely compact objects, where the modified dynamics and the new pressure gradients should play an important role. Since the differential equations have the same degree as those of GR, we do not expect new solutions which may depend on free parameters, as it happens in other types of modified theories which introduce higher-order equations. Rather, the solutions of our set of equations must represent deformations of those found in GR. In fact, the GR solutions should be recovered smoothly and in a unique way in the limit $R_P\to \infty$. The fact that this family of models leads to bounded density and pressure raises a natural question: can we find static solutions corresponding to objects denser than the black holes of GR? This question is pertinent because in GR there can not be static solutions with $r-2M(r)\leq 0$, since they unavoidably lead to gravitational collapse and the divergence of energy density and curvature scalars. The fact that $\rho$ and $P$ are bounded in this theory suggests that such static solutions could exist. Exploring this possibility will be the subject of future research.  \\

This work has been supported by the Spanish grants FIS2008-06078-C03-02, FIS2011-29813-C02-02, the Consolider Program CPAN (CSD2007-00042), and the JAE-doc program of the Spanish Research Council (CSIC). H.S-A. thanks the Department of Theoretical Physics of the University of Valencia for their kind hospitality during the elaboration of this work. The authors thank D. Rubiera-Garcia for useful comments and suggestions.

\end{document}